\documentclass[aps,prl,reprint]{revtex4-2}
\usepackage{blindtext}
\usepackage{graphicx}
\usepackage{subcaption}
\usepackage{amssymb}
\usepackage{xcolor}
\usepackage{graphicx}
\usepackage{floatrow}
\usepackage[fleqn]{amsmath}
\usepackage[font=small,skip=3pt]{caption}
\usepackage{makecell}
\usepackage{hyperref} 
\usepackage{cleveref}
\usepackage{url}
\begin{document}
\title{Doubling Fusion Power with Volumetric Optimization in Magnetic Confinement Fusion Devices}
\author{J. F. Parisi$^{1}$}
\email{jparisi@pppl.gov}
\author{J. W. Berkery$^1$}
\author{A. Sladkomedova$^2$}
\author{S. Guizzo$^3$}
\author{M. R. Hardman$^2$}
\author{J. R. Ball$^4$}
\author{A. O. Nelson$^3$}
\author{S. M. Kaye$^1$}
\author{M. Anastopoulos-Tzanis$^2$}
\author{S. A. M. McNamara$^2$}
\author{J. Dominski$^1$}
\author{S. Janhunen$^2$}
\author{M. Romanelli$^2$}
\author{D. Dickinson$^5$}
\author{A. Diallo$^1$}
\author{A. Dnestrovskii$^2$}
\author{W. Guttenfelder$^{6,1}$}
\author{C. Hansen$^3$}
\author{O. Myatra$^7$}
\author{H. R. Wilson$^8$}
\affiliation{$^1$Princeton Plasma Physics Laboratory, Princeton University, Princeton, NJ, USA}
\affiliation{$^2$Tokamak Energy Ltd., 173 Brook Drive, Milton Park, Oxfordshire, UK}
\affiliation{$^3$Department of Applied Physics \& Applied Mathematics, Columbia University, New York, NY,  USA}
\affiliation{$^4$Ecole Polytechnique Federale de Lausanne, Swiss Plasma Center, Lausanne, Switzerland}
\affiliation{$^5$York Plasma Institute, Department of Physics, University of York, Heslington, York, UK}
\affiliation{$^6$Type One Energy, 8383 Greenway Boulevard, Middleton, WI, USA}
\affiliation{$^7$United Kingdom Atomic Energy Authority, Culham Science Centre, Abingdon, UK}
\affiliation{$^8$Oak Ridge National Laboratory, Oak Ridge, TN, USA}
\begin{abstract}
{A technique, volumetric power optimization, is presented for enhancing the power output of magnetic confinement fusion devices. Applied to a tokamak, this approach involves shifting the burning plasma region to a larger plasma volume while introducing minimal perturbations to the plasma boundary shape. This edge perturbation -- squareness -- is analogous to pinching and stretching the edge boundary. Stability calculations confirm that this edge alteration is compatible with maintaining plasma stability. This optimization method for optimizing fusion power output could improve the performance of magnetic confinement fusion power plants.}
\end{abstract}

\maketitle

\setlength{\parskip}{0mm}
\setlength{\textfloatsep}{5pt}

\setlength{\belowdisplayskip}{6pt} \setlength{\belowdisplayshortskip}{6pt}
\setlength{\abovedisplayskip}{6pt} \setlength{\abovedisplayshortskip}{6pt}

\textit{Introduction.} -- {The variation of fusion power in a power plant is a complex issue, and is one of the primary, if not the primary, cost drivers of a fusion power plant \cite{Wade2021}. In this article, we introduce a technique for optimizing fusion power through the volumetric distribution of the magnetic equilibrium \cite{Shafranov1957,Grad1958}. This approach substantially increases the power output of magnetic confinement systems while enabling variable-power operation, addressing challenges in the practical deployment of fusion energy.}

{The relationship between total thermal fusion power and net power output in fusion power plants is highly nonlinear: increases in thermal power lead to disproportionately large increases in net power output due to recirculating power to plant systems \cite{Menard2016,Buttery2021,Wurzel2022}. As such, strategies that increase total fusion power without raising recirculating power are highly desirable. Furthermore, as electric grids increasingly integrate variable renewable energy sources, the ability of baseload power plants to provide variable power output becomes increasingly valuable \cite{Lund2015,Levin2015,Heptonstall2021}. This trend is reflected in modern fission reactor designs, which offer variable output options despite their historical reliance on fixed operation \cite{Jenkins2018,Cany2018,Dong2021,OECD2021}.}

{In this work, we focus on tokamaks, the leading candidate for commercial fusion energy. Tokamaks confine deuterium-tritium (DT) plasma within a magnetic cage. We show that small adjustments to the magnetic cage shape can significantly increase total fusion power. This capability not only enhances the flexibility of tokamaks but could also substantially reduce the capital cost of fusion power plants \cite{Wade2021}. While various methods for varying power output have been proposed—such as adjusting the fuel mix \cite{Boozer2021,Maslov2023,Parisi_2024d}, density profiles \cite{Wilson2024}, or total plasma volume—each presents significant physics and engineering challenges. For example, increasing fueling rates can impact plasma performance in several ways: exceeding density limits can trigger instabilities \cite{Greenwald1988,Gates2012,Giacomin2022}, modifying the bootstrap current profiles can alter confinement properties \cite{Galeev1971,Peeters2000,Belli2008b}, and reshaping heating and fueling deposition profiles can affect energy distribution \cite{Baylor1992,Garzotti2011,Wolf2018}. Similarly, adjusting the DT fuel mix \cite{Boozer2021,Parisi2024e} can degrade plasma performance due to isotope effects, which influence transport and stability properties \cite{Bessenrodt1993,Xu2013,Garcia2016,Maggi2018}. The approach we present offers a promising alternative solution with lower complexity.}

{We demonstrate the potential of volumetric optimization in an example with plasma squareness. By varying the plasma edge squareness, $\zeta_0$, we demonstrate simultaneous achievement of high maximum fusion power and flexibility to control the fusion burn with minimal disruption to the overall plasma configuration. Increasing $\zeta_0$ redistributes plasma volume to regions of higher power density, leading to significant increases in fusion power. This approach is operationally advantageous as it allows key plasma parameters, such as elongation and triangularity, to remain unchanged while squareness is adjusted. Although this work focuses on plasma squareness, it represents just one application of volumetric optimization.}

\begin{figure}[bt]
    \centering
    \includegraphics[width=0.65\textwidth]{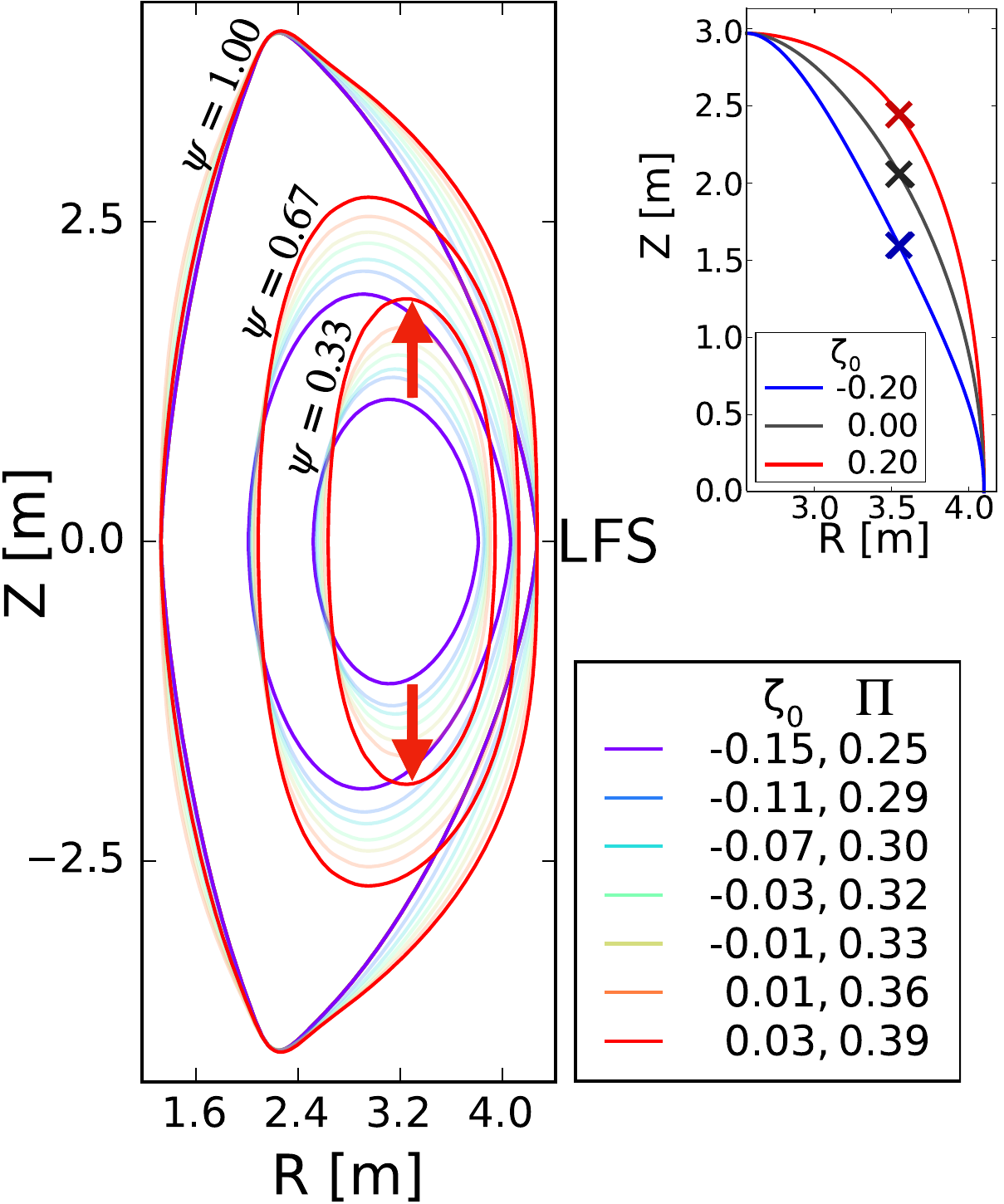}
    \caption{Main: three flux-surfaces $\psi = [0.33, 0.67, 1.0]$ with varying edge squareness $\zeta_0$ and $\Pi$. Red arrows show stretching of inner flux-surfaces by increasing $\zeta_0$. Inset: outer flux-surfaces for three $\zeta_0$ values.}
    \label{fig:fluxsurfshapes}
\end{figure}

{\textit{Volumetric Power Optimization.}} -- Magnetic confinement schemes must confine plasma at sufficient pressure for a high fusion power density $p_{\mathrm{f}}$. However, high $p_{\mathrm{f}}$ is insufficient for substantial fusion power; it must also {persist} over a large fraction of the total plasma volume $V_{\mathrm{tot}}$. The largest contributions to the total fusion power
\begin{equation}
P_{\mathrm{f}} = \int p_{\mathrm{f}} dV,
\label{eq:fusion_totalP}
\end{equation} 
come from the burn volume
\begin{equation}
V_{\mathrm{burn}} \equiv \int_{\mathrm{burn} } d V,
\end{equation}
the volume over which $p_{\mathrm{f}}$ exceeds a value $p_{\mathrm{f,c}}$ required for substantial fusion power. We use $p_{\mathrm{f,c}} = 1$ MW/m$^3$, which is justified later. In this work, we control $P_{\mathrm{f}}$ by changing the fraction of the total plasma volume $V_{\mathrm{tot}}$ packed into $V_{\mathrm{burn}}$, given by the packing {number}
\begin{equation}
\Pi \equiv V_{\mathrm{burn}}/V_{\mathrm{tot}}.
\label{eq:packingfraction}
\end{equation}
Plasmas with $\Pi = 1$ are volume-efficient, packing the full volume into power dense regions. In contrast, $\Pi = 0$ is volume-inefficient since $V_{\mathrm{burn}} = 0$. Thus at fixed $V_{\mathrm{tot}}$,
\begin{equation}
P_{\mathrm{f}} \sim \Pi \; \langle p_{\mathrm{f}} \rangle_{\mathrm{burn}},
\label{eq:fusion_total_Xi}
\end{equation}
where the burn average $\langle \cdot \rangle_{\mathrm{burn}}$ is
\begin{equation}
\langle p_{\mathrm{f}} \rangle_{\mathrm{burn}} \equiv (1/V_{\mathrm{burn}}) \int_{\mathrm{burn}} p_{\mathrm{f}} dV.
\label{eq:core_vol_av}
\end{equation}
The key idea of this work is that {total fusion power can be significantly increased by maximizing $\Pi$ in \cref{eq:packingfraction}, {corresponding to an efficient use of the plasma volume}. In this paper, we demonstrate how to vary $\Pi$ using plasma squareness.} Adjusting the plasma edge squareness \cite{Turnbull1999},
\begin{equation}
\zeta_0 = \arcsin \left( Z_m / \kappa_0 a \right) - \pi/4,
\end{equation}
changes $\Pi$ and thus $P_{\mathrm{f}}$, while $\langle p_{\mathrm{f}} \rangle_{\mathrm{burn}}$ \textcolor{black}{is approximately constant according to infinite-n ballooning stability \cite{Connor1979,Greene1981,Gaur2023}, which we explain in a later section}. Here, $Z_m$ is the $Z$ value indicated by crosses in \cref{fig:fluxsurfshapes}'s inset and $\kappa_0 = h/2a$ is the edge plasma elongation where $h$ is the maximum plasma height and $a$ is the minor radius. By maximizing $\Pi$ in \cref{eq:fusion_total_Xi} using squareness, the fusion power increases significantly.

Dedicated studies of plasma squareness are limited compared with other shape parameters, but benefits to increased $\zeta_0$ have been identified. Squareness control is established \cite{Turnbull1999,Gates2006,Kolemen2011,Ariola2016,Degrave2022} and {less disruptive} than other shape parameters because plasma x-points and maximum width and height can be fixed. Ion-scale turbulence simulations found that higher $\zeta_0$ improves heat confinement \cite{Joiner2010}, and DIII-D and MAST-U experiments show $\zeta_0$ values typically allow higher edge {plasma} pressure \cite{Ferron2000,Lao2001,Leonard2007,Nelson2022,Imada2024} and give improved core confinement \cite{Holcomb2009}. Spherical tokamak design studies found positive moderate $\zeta_0$ stabilized kink modes, allowing a 10\% increase in core plasma pressure \cite{Mau1999,Turnbull1999,Jardin2003}. This benefit was discounted because of the required increase in poloidal field coil current \cite{Jardin2003,Bromberg2003}, but the benefits outlined in this work could change that tradeoff assessment. Furthermore, the poloidal coil currents required for the equilibria in this study are probably easier to handle since we examine significantly lower $\zeta_0$ values. It should be noted that changes to parameters like $\kappa_0$ or $\zeta_0$ may require changes in other plasma parameters, such as internal inductance \cite{Menard2016}. The work in this study is based on a representative burning spherical tokamak \cite{Peng1986} design point, {referred} to as GST. The main GST parameters are on-axis toroidal field $B_{T,0} = 5$T, plasma current $I_p = 9.8$MA, major radius $R_0 = 3.4$m, $a = 1.5$m, $\kappa_0 = 2.73$, and {triangularity} $\delta_0 = 0.43$.

\begin{figure}[!tb]
    \begin{subfigure}[t]{0.31\textwidth}
    \centering
    \includegraphics[width=1.1\textwidth]{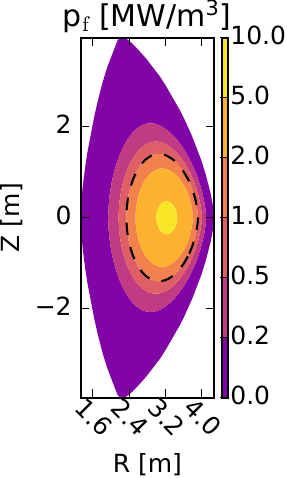}
    \caption{$\zeta_0 = -0.15$}
    \end{subfigure}
    ~
    \begin{subfigure}[t]{0.31\textwidth}
    \centering
    \includegraphics[width=1.1\textwidth]{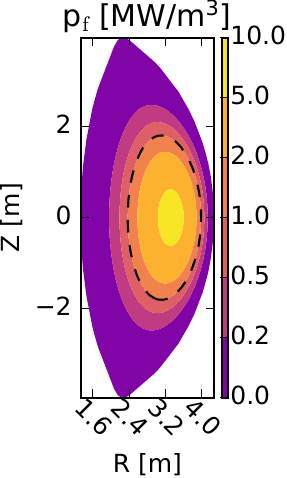}
    \caption{\textcolor{black}{$\zeta_0 = -0.03$}}
    \end{subfigure}
    ~
    \begin{subfigure}[t]{0.31\textwidth}
    \centering
    \includegraphics[width=1.1\textwidth]{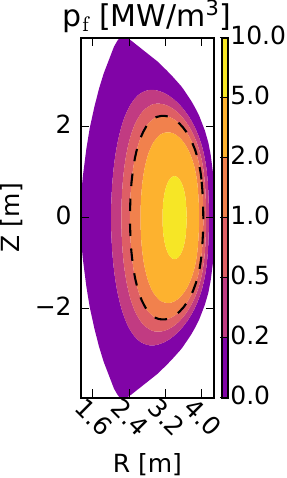}
    \caption{\textcolor{black}{$\zeta_0 = 0.03$}}
    \end{subfigure}
    ~
    \begin{subfigure}[t]{0.31\textwidth}
    \centering
    \includegraphics[width=1.0\textwidth]{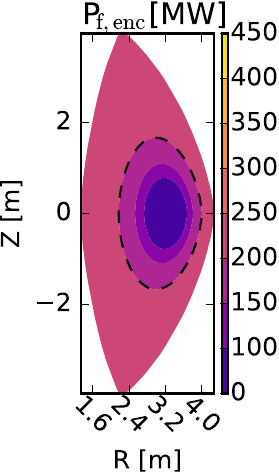}
    \caption{$\zeta_0 = -0.15$}
    \end{subfigure}
    ~
    \begin{subfigure}[t]{0.31\textwidth}
    \centering
    \includegraphics[width=1.0\textwidth]{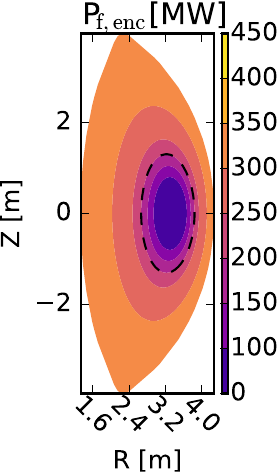}
    \caption{$\zeta_0 = -0.03$}
    \end{subfigure}
    ~
    \begin{subfigure}[t]{0.31\textwidth}
    \centering
    \includegraphics[width=1.0\textwidth]{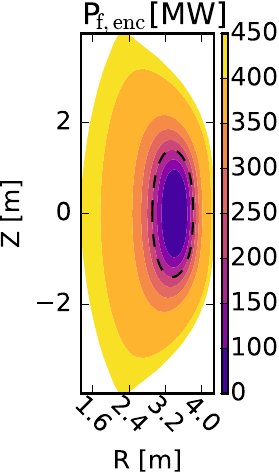}
    \caption{\textcolor{black}{$\zeta_0 = 0.03$}}
    \end{subfigure}
    \caption{Burn control scheme using plasma squareness for GST: Fusion power density (a-c) and enclosed fusion power (d-f) in increasingly square plasmas. The $p_{\mathrm{f} } = 1$ MW/m$^{3}$ and $P_{\mathrm{f} } = 200$ MW surfaces are indicated with a dashed black contour. The total fusion power is 215 MW, 311 MW, and 403 MW for $\zeta = -0.15, -0.03$, and $0.03$.}
    \label{fig:power_density}
\end{figure}

\textit{Squareness.} -- Increasing squareness of the plasma edge stretches the magnetic surfaces in the plasma core. An equilibrium with $\zeta_0 = 0.03$ has core flux-surfaces that are twice as elongated as a lower $\zeta_0$ equilibrium with $\zeta_0 = -0.15$. This is illustrated in \cref{fig:fluxsurfshapes} -- we plot surfaces of constant {normalized} poloidal flux $\psi$ with different $\zeta_0$ and almost constant plasma volume. Comparison of the inner $\psi = 0.33$ flux surfaces for $\zeta_0 = -0.15$ and $\zeta_0 = 0.03$ shows that higher $\zeta_0$ elongates the plasma core.

Increased core elongation more efficiently uses the core volume for fusion power by increasing $V_{\mathrm{burn}}$ and hence $\Pi \sim \langle \kappa \rangle_{\mathrm{burn}}$. The total power is therefore determined mainly by $p_{\mathrm{f}}$ and $\kappa$ in the burn region, 
\begin{equation}
P_f \sim \langle \kappa \rangle_{\mathrm{burn}} \; \langle p_{\mathrm{f}} \rangle_{\mathrm{burn}},
\end{equation}
for fixed $V_{\mathrm{tot} }$. Therefore, $\zeta_0$, mainly through $\langle \kappa \rangle_{\mathrm{burn}}$, determines $\Pi$ and thus the fusion burn.

The physical mechanism for $\zeta_0$ increasing $\Pi$ exploits plasma properties at the low-field-side (LFS), indicated in \cref{fig:fluxsurfshapes}. With increasing plasma squareness, $\Pi$ increases from 0.25 to 0.39, shown in \cref{fig:fluxsurfshapes}. The strong poloidal field at the LFS causes LFS flux-surfaces to be closely spaced, measured by large $|d \psi / d r|_{\mathrm{LFS}}$ values, where $r$ is the plasma minor radial coordinate. When $\zeta_0$ increases, flux-surfaces {in the plasma core} are stretched vertically in order to keep $|d \psi / d r|_{\mathrm{LFS}}$ large, indicated by red arrows in \cref{fig:fluxsurfshapes}. This effect increases for spherical tokamaks where the LFS poloidal field is particularly strong \cite{Peng1986} and for high Shafranov shift \cite{Shafranov1962}, which both increase $|d \psi / d r|_{\mathrm{LFS}}$.

\begin{figure}[tb]
    \begin{subfigure}[t]{0.47\textwidth}
    \centering
    \includegraphics[width=1.03\textwidth]{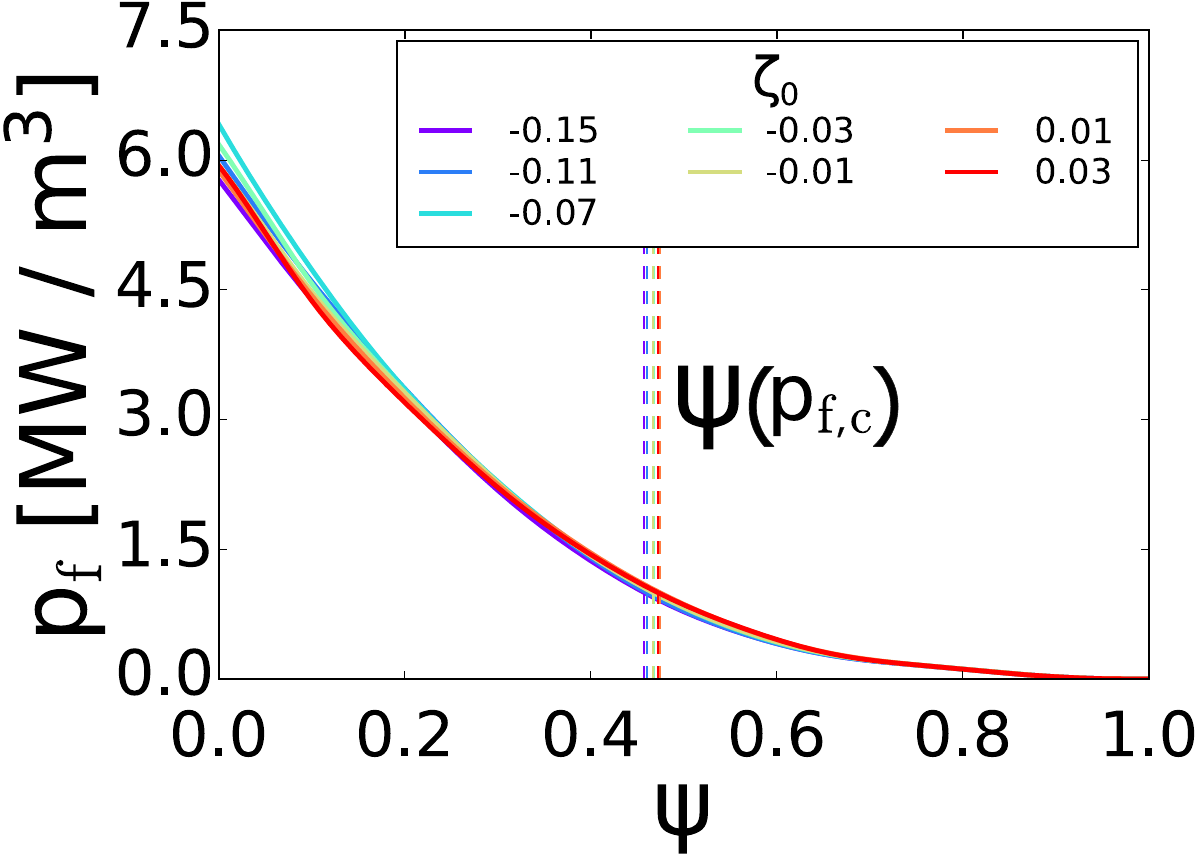}
    \caption{\textcolor{black}{Fusion power density}}
    \end{subfigure}
     ~
    \begin{subfigure}[t]{0.46\textwidth}
    \centering
    \includegraphics[width=1.0\textwidth]{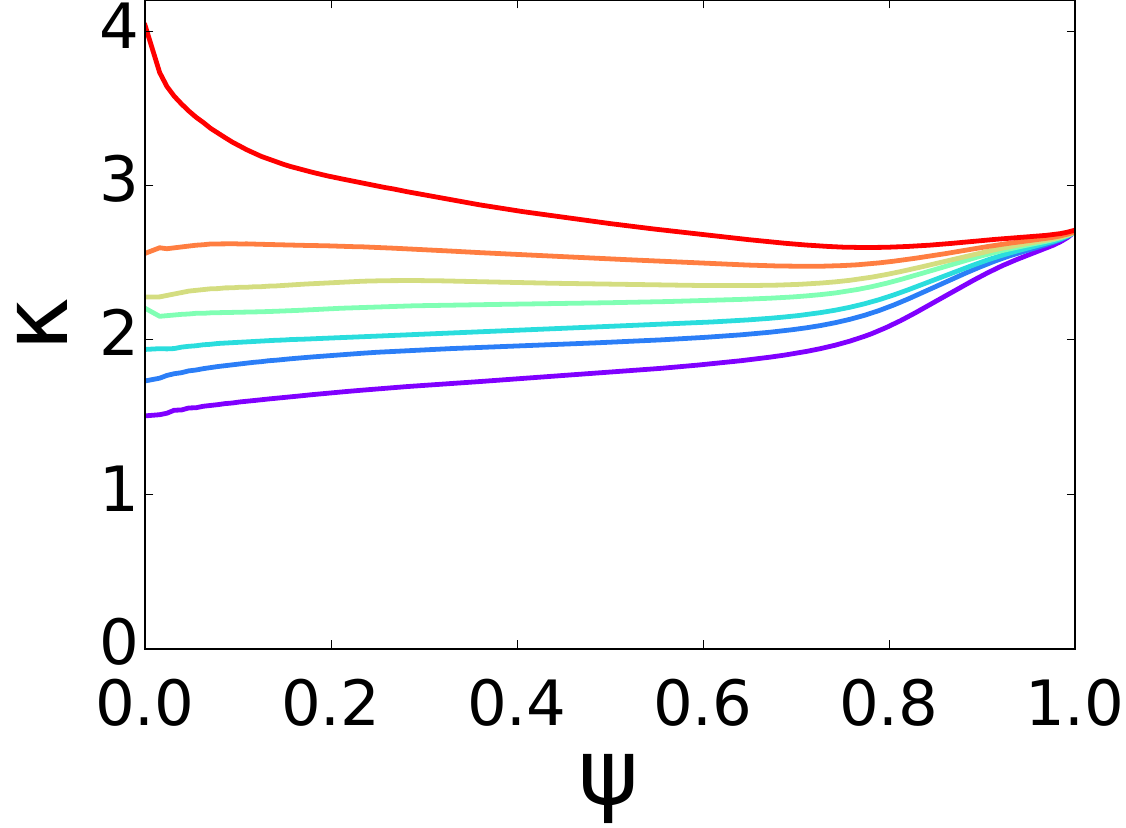}
    \caption{\textcolor{black}{Elongation}}
    \end{subfigure}
    ~
    \centering
    \begin{subfigure}[t]{0.47\textwidth}
    \centering
    \includegraphics[width=1.03\textwidth]{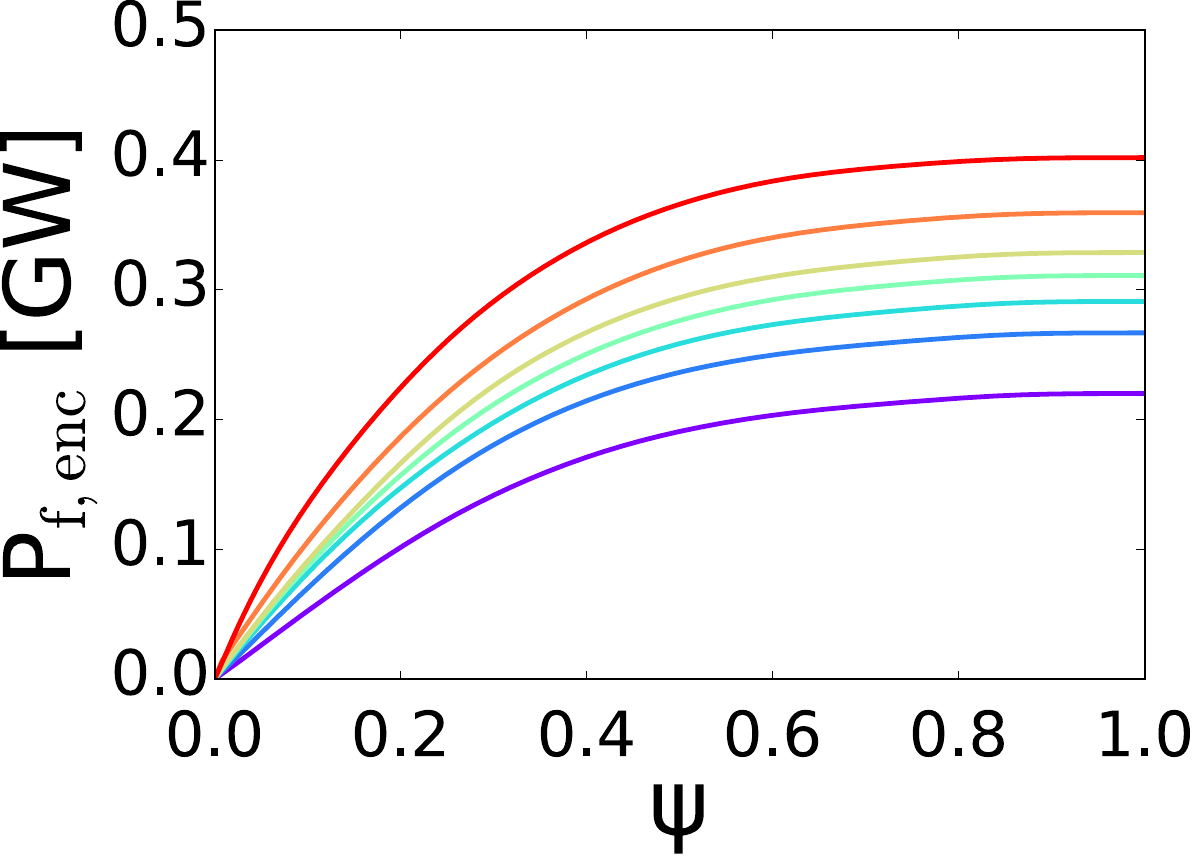}
    \caption{\textcolor{black}{Enclosed fusion power}}
    \end{subfigure}
     ~
    \begin{subfigure}[t]{0.47\textwidth}
    \centering
    \includegraphics[width=1.03\textwidth]{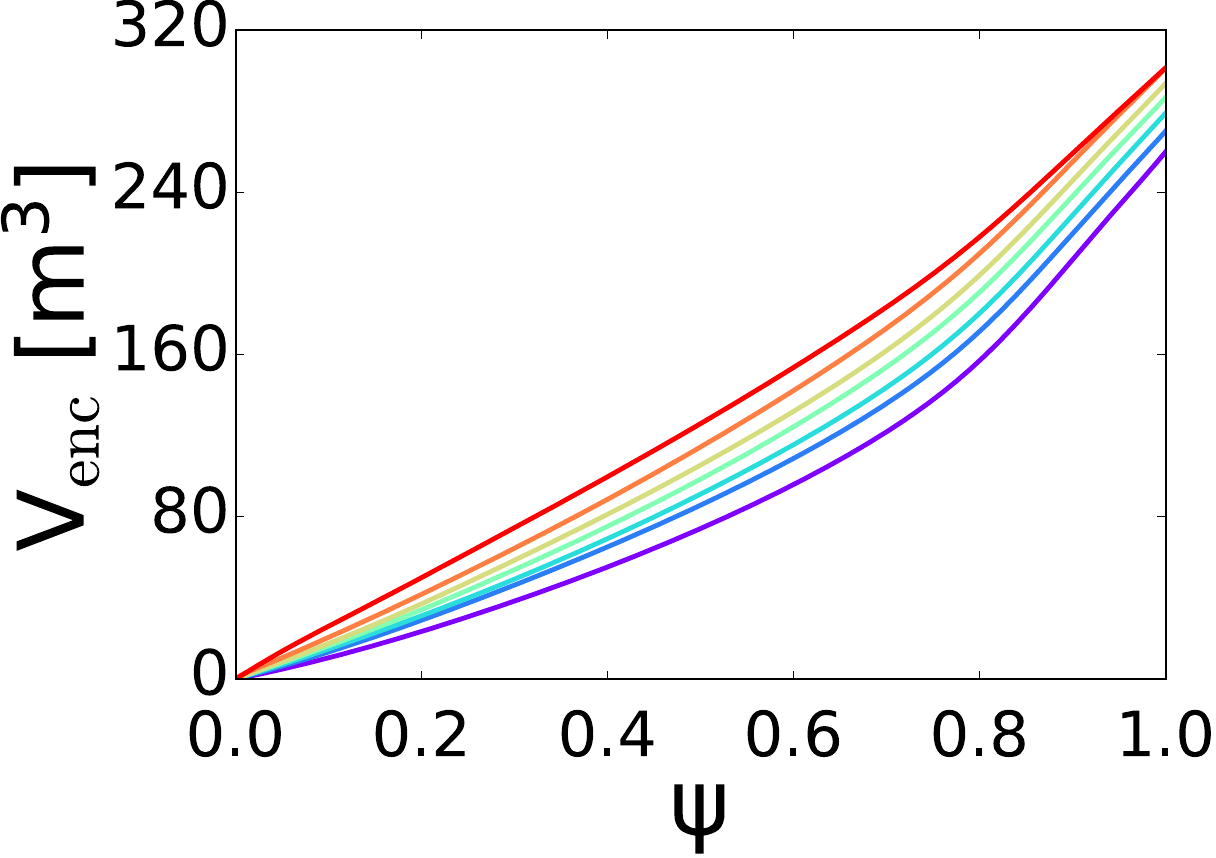}
    \caption{\textcolor{black}{Enclosed plasma volume}}
    \end{subfigure}
     ~
    \begin{subfigure}[t]{0.47\textwidth}
    \centering
    \includegraphics[width=1.03\textwidth]{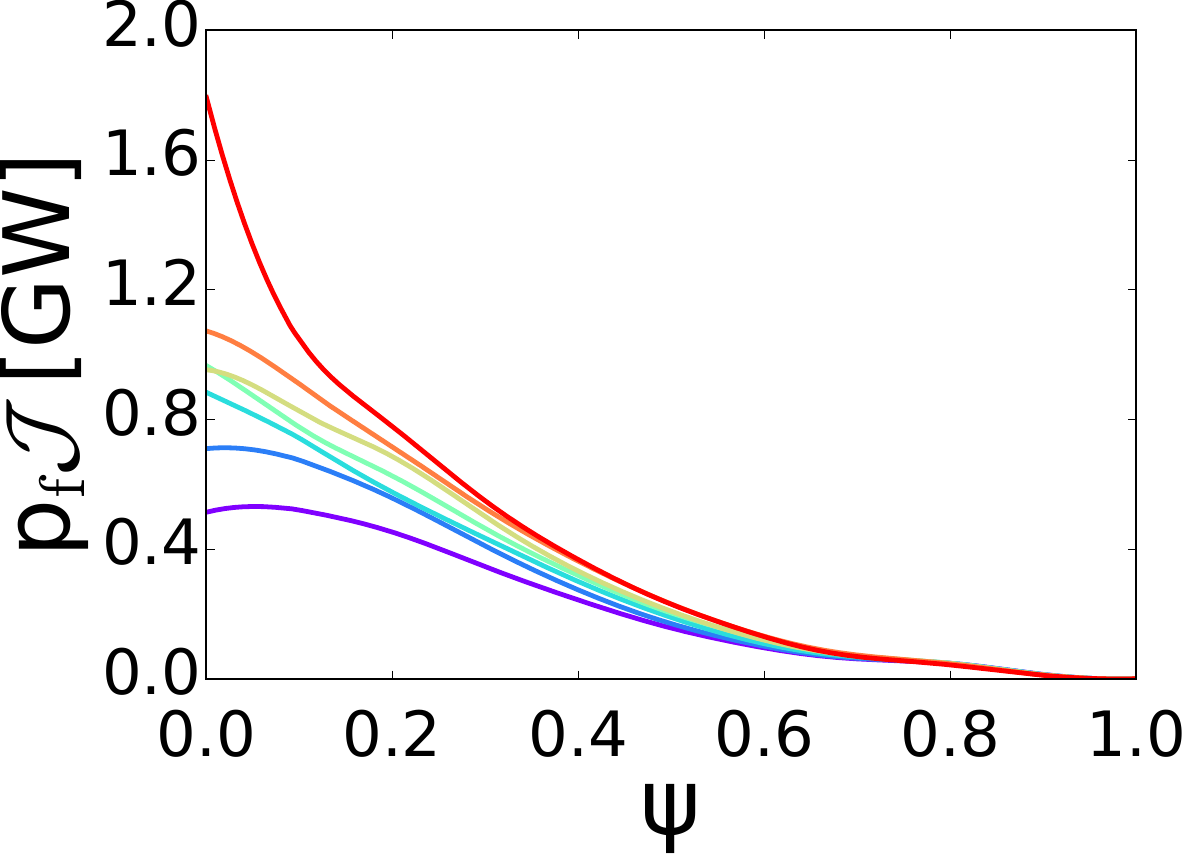}
    \caption{\textcolor{black}{Enclosed power derivative}}
    \end{subfigure}
     ~
    \begin{subfigure}[t]{0.47\textwidth}
    \centering
    \includegraphics[width=1.03\textwidth]{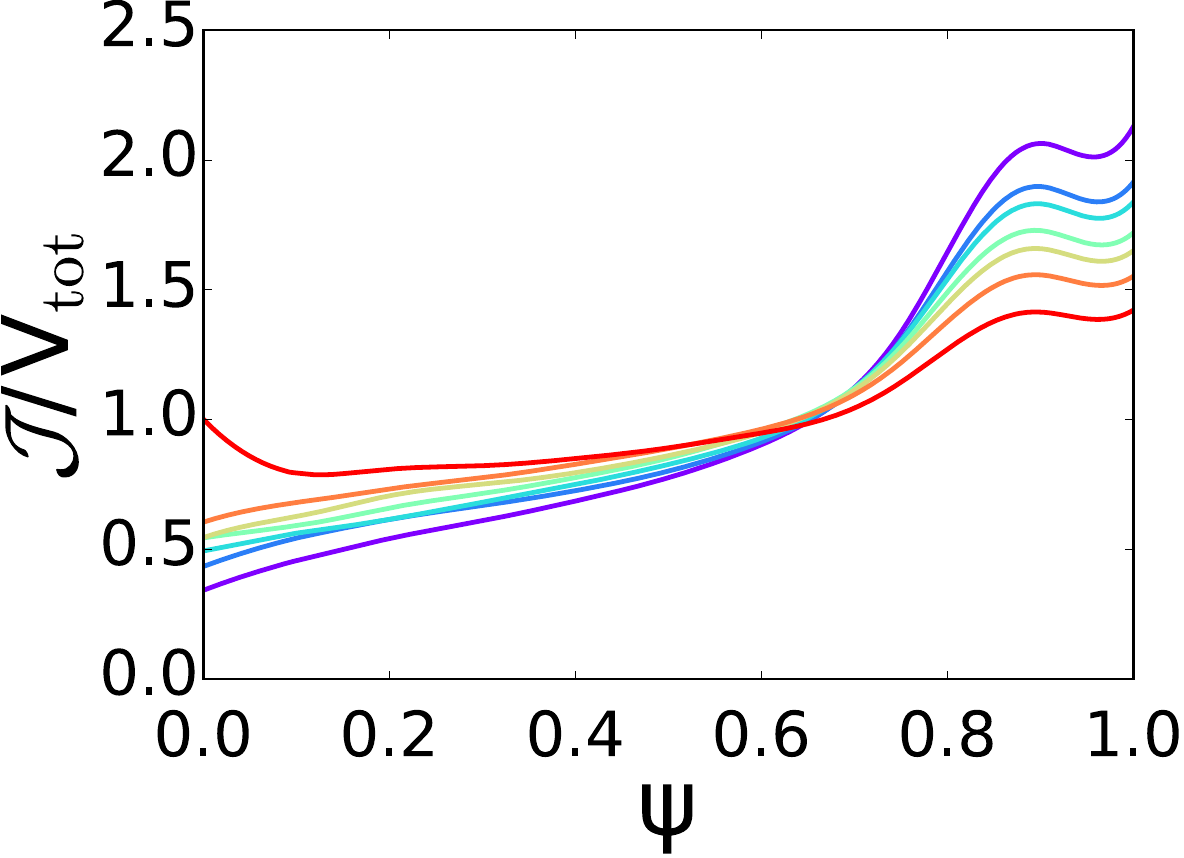}
    \caption{Stretch}
    \end{subfigure}
    \caption{GST radial profiles of power density (a), elongation (b), enclosed power (c), enclosed volume (d) and enclosed power derivative (\cref{eq:dP_contributions}) (e), and Stretch (\cref{eq:stretch}) {(f)}.}
    \label{fig:power_stats}
\end{figure}

The total power $P_{\mathrm{f}}$ doubles from the minimum to maximum $\zeta_0$ values, from $P_{\mathrm{f} } = 215$ MW to $403$ MW. The corresponding $p_{\mathrm{f}}$ is plotted for plasmas with three $\zeta_0$ values in \cref{fig:power_density}(a)-(c). The highest $\zeta_0$ equilibrium has $p_{\mathrm{f}}$ surfaces that are much more elongated (\cref{fig:power_stats}(b)) and it therefore has a high $\Pi$ value ($\Pi = 0.39$ compared with $\Pi = 0.25, 0.32$ for $\zeta_0 = -0.15, -0.03$). Thus, higher $\zeta_0$ equilibria efficiently distribute volume to regions of high $p_{\mathrm{f}} (\psi)$, which increases $P_{\mathrm{f}}$, despite similar $p_{\mathrm{f}} (\psi)$ profiles (\cref{fig:power_stats}(a)). For example, for $\zeta_0 = 0.03$ the flux-surface $\psi = 0.2$ has an enclosed power
\begin{equation}
P_{\mathrm{f, enc}} (\psi) \equiv \int_0^{V(\psi)} p_{f} dV'
\end{equation}
of 231 MW but $\zeta_0 = -0.15$ has only 98 MW (\cref{fig:power_density}(d)-(f) and \cref{fig:power_stats}(c)). In \cref{fig:power_stats}(a), vertical lines indicate the surface $\psi_{\mathrm{burn}}$ where $p_{\mathrm{f,c}} = 1$ MW/m$^3$; for all equilibria, $P_{\mathrm{f,enc}}(\psi_{\mathrm{burn}})/P_{\mathrm{f}} > 90\%$, indicating a good choice of $p_{\mathrm{f,c} }$.

\begin{figure}[tb]
\ffigbox[7.8cm]{%
\begin{subfloatrow}
  \hsize0.5\hsize
  \vbox to 7.35cm{
  \ffigbox[\FBwidth]
    {\caption{Deuterium density}}
    {\includegraphics[width=0.5\textwidth]{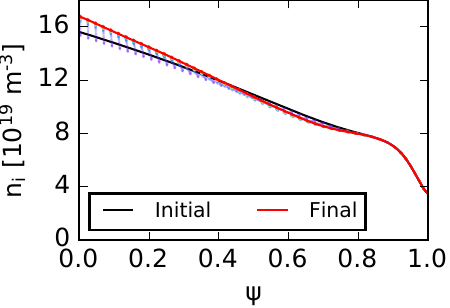}}\vss
  \ffigbox[\FBwidth]
    {\caption{Deuterium temperature}}
    {\includegraphics[width=0.5\textwidth]{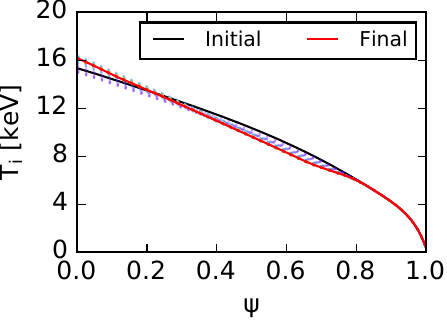}}
  }
\end{subfloatrow}\hspace*{\columnsep}
\begin{subfloatrow}
  \ffigbox[\FBwidth][]
    {\caption{Flux surfaces $\psi = [0.33, 0.67, 1.00]$}}
    {\includegraphics[width=0.4\textwidth]{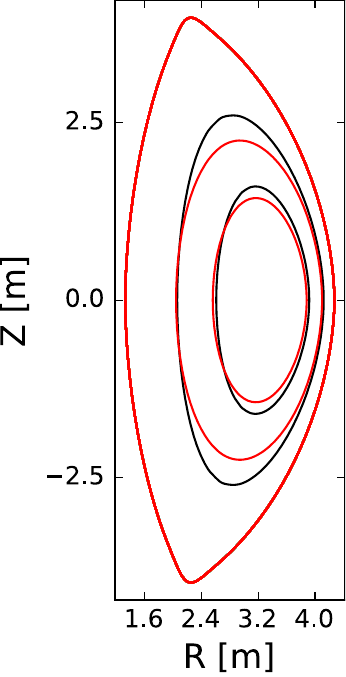}}
\end{subfloatrow}
}{
\caption{Initial and final ion density and temperature for GST ((a),(b)), and equilibria (c) for $\zeta_0 = -0.01$. In (a) and (b), dotted lines are intermediate iterations. The final profile is infinite-n ballooning stable.}
\label{fig:ballooning_optimizer}
}
\end{figure}

Higher $\zeta_0$ equilibria enclose more fusion power by growing volume relatively quickly in the burn region but slowly outside. The Jacobian $\mathcal{J} = dV_{\mathrm{enc}} / d\psi$ measures flux-surface growth for enclosed volume $V_{\mathrm{enc} }(\psi) \equiv \int_0^{V(\psi)} d V'$ (\cref{fig:power_stats}(d)). $\mathcal{J}$ enters the derivative of $P_{\mathrm{f, enc}}$,
\begin{equation}
d P_{\mathrm{f,\mathrm{enc}} } / d \psi = p_{\mathrm{f}} \mathcal{J},
\label{eq:dP_contributions}
\end{equation}
so larger $\mathcal{J}$ indicates faster flux-surface volume growth, contributing more to $P_{\mathrm{f}}$. Average contributions to $P_{\mathrm{f} }$ in the burn region are double for the highest $\zeta_0$ than the lowest $\zeta_0$, shown by $p_{\mathrm{f}} \mathcal{J}$ in \cref{fig:power_stats}(e). This explains how the highest $\zeta_0$ has double the $P_{\mathrm{f}}$ of the lowest $\zeta_0$. The Stretch $S$ is the normalized rate of volume growth,
\begin{equation}
S = \mathcal{J} / V_{\mathrm{tot} },
\label{eq:stretch}
\end{equation}
satisfying $\int_0^{1} S d\psi = 1$. Surfaces with $S > 1$ increase volume faster than the average flux-surface and those with $S < 1$ increase volume slower than average. Because $S \sim (\kappa/\kappa_0) (r/a) (B_{\mathrm{p},0} / B_\mathrm{p})$, where $B_\mathrm{p,0}$ is the edge poloidal field, we expect $S \ll 1$ near the core but $S \gg 1$ at the edge. This is true for low $\zeta_0$ equilibria, shown in \cref{fig:power_stats}(f), but for high $\zeta_0$ equilibria, $S \approx 1$ in $V_{\mathrm{burn}}$, and increases slowly near the edge. Therefore, high $\zeta_0$ equilibria expand plasma volume relatively quickly in the burn region, whereas low $\zeta_0$ equilibria expand plasma volume relatively quickly in the edge where $p_{\mathrm{f}}$ is low.

\textit{Equilibrium Generation.} -- While this paper's focus is on new geometry insights and not integrated modeling, we ensure that the equilibria presented here clear a minimum threshold for viability: infinite-n ballooning stability \cite{Connor1979,Greene1981,Gaur2023}{, which is an approximation for kinetic-ballooning-mode stability \cite{Tang1980,Aleynikova2018}.} We found that changing the equilibrium magnetic geometry by varying $\zeta_0$ but keeping the profiles fixed caused some equilibria to be ballooning unstable. Therefore, for a fair comparison across $\zeta_0$ values, we use an iterative scheme that generates equilibria close to the infinite-n ballooning stability boundary. For each iteration, at 10 radial locations the density and temperature gradients for thermal species are brought to the ballooning stability boundary. The fixed-boundary equilibrium is recalculated using CHEASE \cite{Lutjens1996} according to the new pressure and current profiles (with consistent bootstrap current \cite{Sauter1999,Redl2021}), subject to keeping the total plasma current $I_p$ fixed. Increasing $\zeta_0$ also increases the bootstrap fraction $f_{\mathrm{bs}}$ from $f_{\mathrm{bs}} = 0.72$ to $f_{\mathrm{bs}} = 0.96$ for the lowest to the highest $\zeta_0$. The strong increase in $f_{\mathrm{bs}}$ results from increased core elongation at higher $\zeta_0$ \cite{Menard1997}. The equilibria have five thermal species -- deuterium, tritium, lithium, helium, and electrons -- and a fast helium population. In \cref{fig:ballooning_optimizer}(a) and (b), we plot the deuterium temperature and density at each iteration step, and in \cref{fig:ballooning_optimizer}(c) flux-surfaces for the initial and final equilibria. To simplify analysis, we present results with fixed pressure in the pedestal edge region. Because of the effect of shape on pedestal performance \cite{Osborne2000,Ferron2000,Snyder2015,Parisi_2024b}, we performed sensitivity analysis that showed the trends in this work still held with different pedestal heights.

\begin{figure}[tb]
    \centering
    \includegraphics[width=0.8\textwidth]{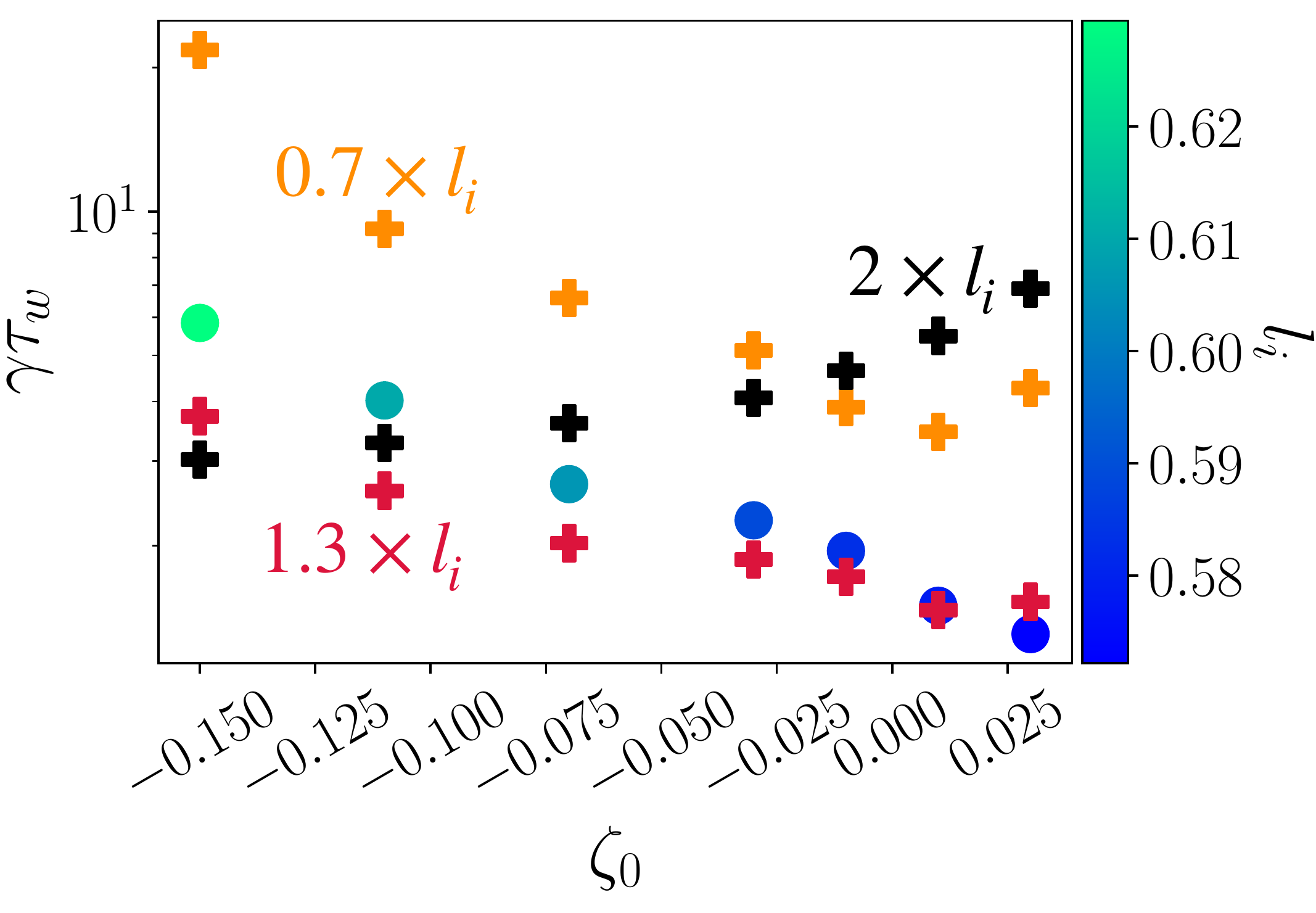}
    \caption{Vertical stability growth rates $\gamma \tau_\mathrm{w}$ across squareness for GST for nominal $l_i$ (circles) and rescaled $l_i$ (pluses).}
    \label{fig:vertical_stab}
\end{figure}

\textit{Vertical Stability.} -- High plasma elongation can make a plasma more vertically unstable and trigger violent plasma termination events \cite{Haas1975,Lazarus1990,Ward1993,Sweeney2020,Nelson2024,Guizzo2024}. Thus, it is important to determine how $\zeta_0$ and its strong effects on $\langle \kappa \rangle_{\mathrm{burn}}$ (\cref{fig:power_stats}(b)) affect vertical stability (VS). To assess VS, we use \texttt{Tokamaker} \cite{Hansen2024} to find VS growth rates for equilibria with different $\zeta_0$ values. Free-boundary equilibria were generated and the VS growth rate $\gamma \tau_w$ calculated for the $n=0$ mode, where the growth rate $\gamma$ is normalized to the wall resistive time $\tau_\mathrm{w}$ and $n$ is the toroidal mode number. Shown in \cref{fig:vertical_stab}, at the nominal inductance $l_i$ values, $0.57 < l_i < 0.63$, increasing $\zeta_0$ is stabilizing. We rescaled $l_i$ by \textcolor{black}{factors of} 0.7, 1.3 and 2.0, finding that slightly higher $l_i$ was stabilizing, but much higher and lower $l_i$ became destabilizing. We hypothesize $\zeta_0$ stabilization at nominal $l_i$ values is due to a larger Shafranov shift at higher squareness, leading to stronger plasma-wall coupling. At higher $l_i$, the Shafranov shift is comparable for all $\zeta_0$ values, and the effect of increased $\langle \kappa \rangle_{\mathrm{burn}}$ with higher $\zeta_0$ dominates. At lower $l_i$, the core surfaces are much more elongated for all $\zeta_0$ values, driving vertical instability. For each $\zeta_0$ value, the plasma wall was conformal to the plasma boundary shape. While detailed studies are required, these results suggest that the $\zeta_0$ values investigated here are compatible with VS stabilization methods \cite{Lee1999,Humphreys2009,Buxton2018,Pesamosca2022}, which in current machines can control $\tau_\mathrm{w} \lesssim 6$, although control may be somewhat degraded in future devices \cite{Lee2015_two}.

\begin{figure}[tb]
\ffigbox[7.8cm]{%
\begin{subfloatrow}
  \hsize0.6\hsize
  \vbox to 6.35cm{
  \ffigbox[\FBwidth]
    {\caption{}}
    {\includegraphics[width=0.5\textwidth]{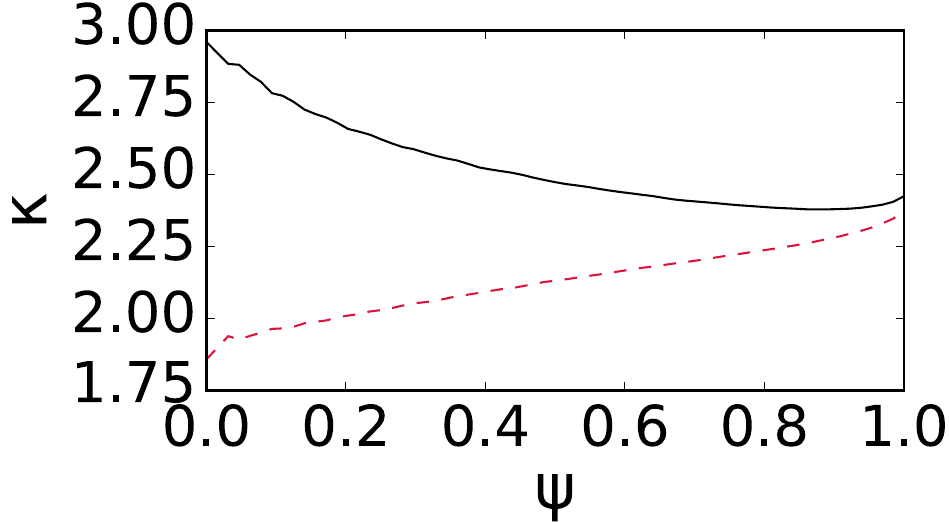}}\vss
  \ffigbox[\FBwidth]
    {\caption{}}%
    {\includegraphics[width=0.5\textwidth]{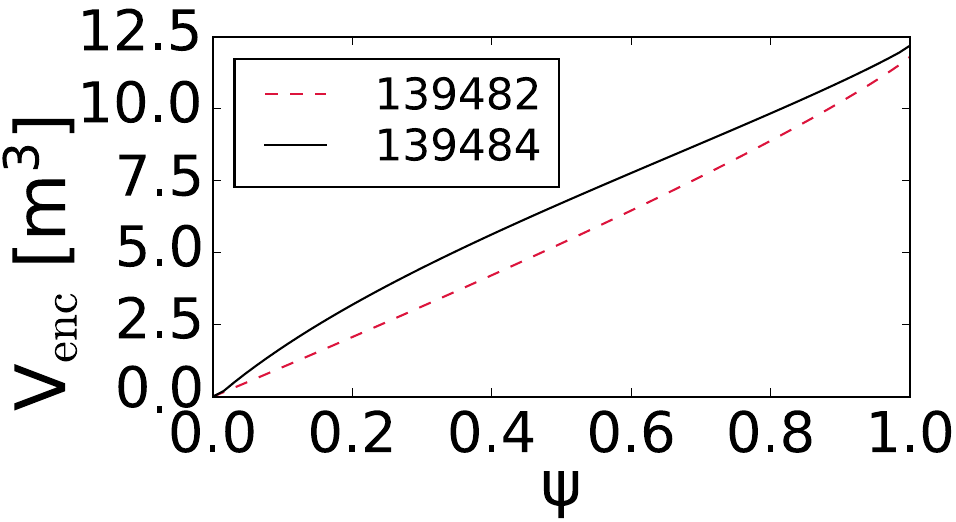}}
  }
\end{subfloatrow}\hspace*{\columnsep}
\begin{subfloatrow}
  \ffigbox[\FBwidth][]
    {\caption{Flux surfaces $\psi = [0.33, 0.67, 1.0]$}}
    {\includegraphics[width=0.4\textwidth]{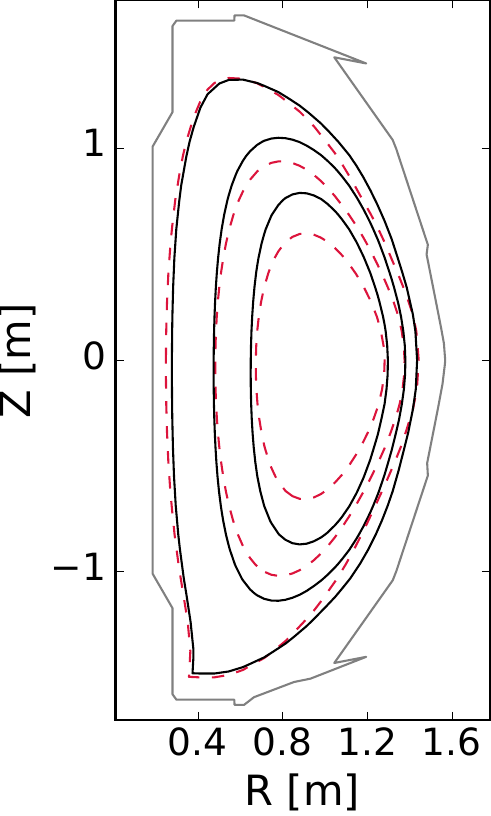}}
\end{subfloatrow}
}{
\caption{Plasma elongation (a), enclosed volume (b), and flux surfaces (c) for two companion NSTX discharges (139482 @547ms, 139484 @556ms) with differing edge squareness ($\zeta = -0.01$ and $\zeta = 0.10$).}
\label{fig:nstx_experiments}
}
\end{figure}

\textit{Experimental Equilibria.} -- While tokamaks have not yet produced burning plasmas, core volume packing using squareness has been observed on current devices. In \cref{fig:nstx_experiments}, we plot the plasma elongation, enclosed volume, and flux surfaces for two companion National Spherical Torus Experiment (NSTX) discharges \cite{Kolemen2011}, which differ mainly by edge squareness ($\zeta_0 = -0.01$ and $\zeta_0 = 0.10$). For the larger $\zeta_0$ discharge, the elongation increases up to the magnetic axis (\cref{fig:nstx_experiments}(a)) and the enclosed plasma volume is much larger in the core (\cref{fig:nstx_experiments}(b)). Three flux surfaces with $\psi = [0.33, 0.67, 1.00]$ are plotted in \cref{fig:nstx_experiments}(c). While the fusion power is negligible in these NSTX discharges, they show that increased core volume via $\zeta_0$ has been realized on current day experiments. Experimental consideration of this effect could continue in NSTX-Upgrade \cite{Berkery_2024}, which is designed to produce a wide range of squareness shapes \cite{Menard2012}.

\textit{Triangularity Scan.} -- Extending the concept of controlling $\Pi$ and $\langle \kappa \rangle_{\mathrm{burn}}$ with $\zeta_0$, we show that edge triangularity $\delta_0$ further redistributes the volume and hence changes $P_{\mathrm{f}}$. Starting from $\zeta_0 = -0.03$, we reconstruct equilibria with varying $\zeta_0$ and $\delta_0$ for fixed pressure and current profiles. Notably, \textcolor{black}{in contrast to the earlier section,} \textcolor{black}{we have fixed $p_{\mathrm{f}}(\psi)$ for these equilibria}, illustrating a purely geometric effect on $P_{\mathrm{f}}$. 

\begin{figure}[!tb]
    \begin{subfigure}[t]{0.9\textwidth}
    \centering
    \includegraphics[width=1.0\textwidth]{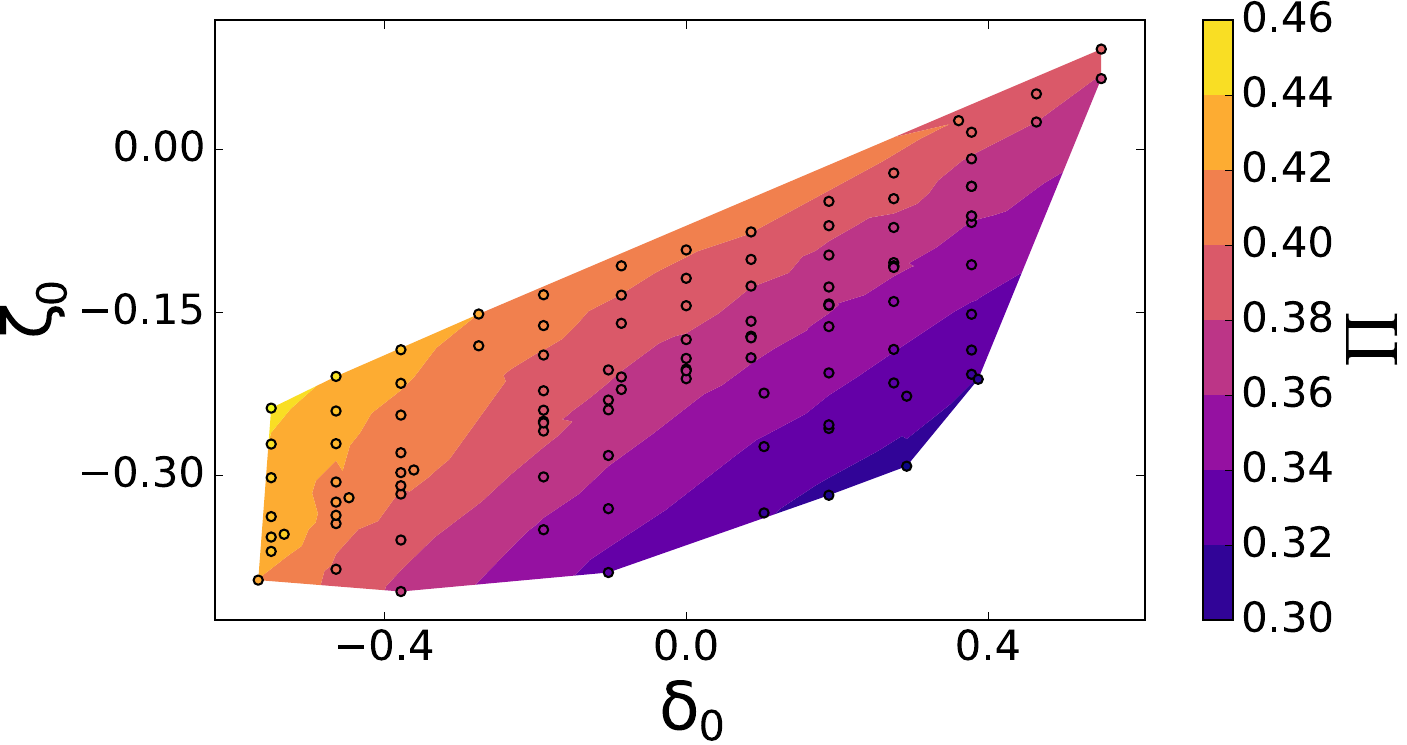}
    \caption{packing {number} $\Pi$ across $\delta_0$ and $\zeta_0$}
    \end{subfigure}
    ~
    \begin{subfigure}[t]{0.31\textwidth}
    \centering
    \includegraphics[width=1.0\textwidth]{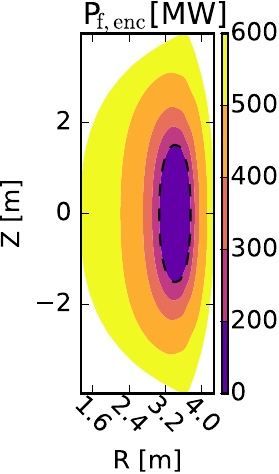}
    \caption{$\delta_0 = -0.5$}
    \end{subfigure}
    ~
    \begin{subfigure}[t]{0.31\textwidth}
    \centering
    \includegraphics[width=1.0\textwidth]{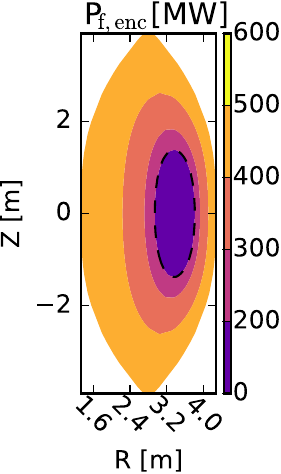}
    \caption{$\delta_0 = 0.0$}
    \end{subfigure}
    ~
    \begin{subfigure}[t]{0.31\textwidth}
    \centering
    \includegraphics[width=1.0\textwidth]{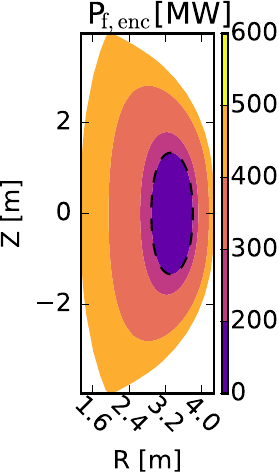}
    \caption{$\delta_0 = 0.5$}
    \end{subfigure}
    \caption{(a) packing {number} $\Pi$ values for triangularity $\delta_0$ and squareness $\zeta_0$ scan for GST at fixed $p_{\mathrm{f}}(\psi)$. (b)-(d): $P_{\mathrm{f,enc}}$ for $\delta_0 = -0.5, 0.0, 0.5$ for $\zeta_0$ with the highest $P_{\mathrm{f}}$.}
    \label{fig:power_density_shapescans}
\end{figure}

Due to strong flux-expansion, negative triangularity (NT) \cite{Austin2019,Nelson2023,Paz-Soldan_2024} gives a higher packing {number} $\Pi$. NT elongates the LFS edge flux-surface, consequently elongating core flux-surfaces. Increasing $\zeta_0$ and reducing $\delta_0$ expands the available core volume for fusion power, shown in \cref{fig:power_density_shapescans} (a). This leads to a higher $P_{\mathrm{f,enc}}$, plotted in \cref{fig:power_density_shapescans}(b)-(d) for three $\delta_0$ values, each with \textcolor{black}{the $\zeta_0$ value yielding the highest $P_{\mathrm{f}}$.} The $\delta_0 = -0.5$ case yields $P_{\mathrm{f} } =$ 520 MW, while $\delta_0 = 0.5$ yields $P_{\mathrm{f} } =$ 420 MW.

From the practical perspective of a plant operator, varying $\delta_0$ is more challenging than $\zeta_0$ \cite{Turnbull1999}. As \cref{fig:power_density_shapescans}(a) shows, while NT increases $\Pi$ at fixed $p_{\mathrm{f}}$, varying $\zeta_0$ for NT does not vary $\Pi$ by nearly as much than at higher $\delta_0$ values. Thus, designers of a variable power tokamak seeking a large $P_{\mathrm{f}}$ range might opt for higher $\delta_0$ at the cost of reduced maximum $P_{\mathrm{f}}$.

{\textit{Generalized volumetric optimization}. -- In this work we focused on increasing $\Pi$ in tokamaks using a combination of Shafranov shift, squareness, and triangularity. There are other ways to achieve high $\Pi$. For example, analytic forms of $d^2V/d\psi^2$ for quasisymmetric stellarators \cite{Landreman2020} could be included in stellarator optimization calculations \cite{Gates2017,Landreman2022,Velasco2023,Jorge2023}. There is some operational flexibility in the current profiles of stellarators \cite{Zarnstorff2001}, which could be used for further optimization. Volumetric optimization could also be achieved in magnetic mirrors by optimizing the field strength along magnetic field lines \cite{Post1987,Bagryansky2015,Endrizzi2023}.}

\textit{Discussion.} -- {This work demonstrates a volumetric power optimization method capable of doubling the fusion power in a tokamak burning plasma with only a 15\% increase in total plasma volume, all while preserving plasma stability, plasma height, and plasma width. This not only enhances the achievable maximum fusion power but also introduces flexibility for variable power output, which could be useful for future energy demands. Furthermore, this optimization framework is adaptable and could be extended to other advanced techniques and alternative fusion confinement concepts beyond those explored here, with broader applications in fusion energy research. This may offer an accelerated route to fusion energy in magnetic confinement devices.}

The concept introduced here of redistributing plasma volume is also a new way to think about tokamak `size.' The two main pathways for modifying `size' in modern tokamaks are the major radius $R$ and the magnetic field strength $B$ \cite{Shimada2007,Sorbom2015,Whyte2016,Zohm2019}. {Total fusion power can increase with $R$ since total plasma volume since scales as $V \sim R^3$. Total fusion power can increase with $B$ since power density scales as $p_f \sim B^4$ at fixed plasma $\beta \sim p / B^2$ where $p$ is the plasma pressure.} In this paper, we have shown a complementary third way, which is the efficiency of the burn volume -- $\Pi$ (see \cref{eq:packingfraction}) -- at roughly fixed $V$ and $B$.

Finally, while we have introduced the concept of {volumetric power optimization}, there are many important physics and engineering areas that we have not analyzed. Fully integrated physics and engineering studies and dedicated experiments are required to determine how best to employ {volumetric power optimization.}

\textit{Acknowledgements.} -- We are grateful to G. W. Hammett, E. J. Paul, and M. C. Zarnstorff for discussions. This work was funded under the INFUSE program – a DOE SC FES public-private partnership –  under contract no. 2706 between Princeton Plasma Physics Laboratory and Tokamak Energy Ltd. This work was supported by the U.S. Department of Energy under contract number DE-SC0022270.

{\textit{Data and Code Availability.} -- Data used in this analysis will be provided upon reasonable request to the first author. Part of the data analysis was performed using the OMFIT integrated modeling framework \cite{OMFIT2015} using the Github projects \texttt{gk\_ped} \cite{Parisi2023a,Parisi_2024} and \texttt{ideal-ballooning-solver} \cite{Gaur2023a}.}

\bibliographystyle{apsrev4-2} %
\bibliography{EverythingPlasmaBib} %

\end{document}